\documentclass{optica-article}
\journal{opticajournal} 
\articletype{ Opt.\ Express}
\usepackage[english]{babel}
\usepackage{siunitx}
\usepackage{nicefrac}
\usepackage{lineno}
\DeclareSIUnit\mbps{Mbps}

\begin{document}
\title{A warm Rydberg atom-based quadrature amplitude-modulated receiver
}

\author{Jan Nowosielski,\authormark{1,2,\textdagger} Marcin Jastrzębski,\authormark{1,2,\textdagger} Pavel Halavach,\authormark{1,2} Karol Łukanowski,\authormark{2} Marcin Jarzyna,\authormark{2,3} Mateusz Mazelanik,\authormark{2} Wojciech Wasilewski,\authormark{1,2} and Michał Parniak\authormark{1,2,*}}

\address{
\authormark{1}Faculty of Physics, University of Warsaw, L. Pasteura 5, 02-093 Warsaw, Poland\\
\authormark{2}Centre for Quantum Optical Technologies, Centre of New Technologies, University of Warsaw, S. Banacha 2c, 02-097 Warsaw, Poland\\
\authormark{3}Department of Optics, Palacky University, 17. listopadu 12, 771 46 Olomouc, Czech Republic\\
}
\address{\authormark{\textdagger} These authors contributed equally}
\email{\authormark{*}mparniak@fuw.edu.pl} 


\begin{abstract*}
Rydberg atoms exhibit both remarkable sensitivity to electromagnetic fields making them promising candidates for revolutionizing field sensors and, unlike conventional antennas, they neither disturb the measured field nor necessitate extensive calibration procedures. In this study, we propose a receiver design for data-modulated signal reception near the 2.4 GHz Wi-Fi frequency band, harnessing the capabilities of warm Rydberg atoms. Our focus lies on exploring various quadrature amplitude modulations and transmission frequencies through heterodyne detection. We offer a comprehensive characterization of our setup, encompassing the atomic response frequency range, attainable electric field amplitudes, and sensitivity, which we estimate to be equal to $\SI{0.50}{\micro\volt\per\cm\per\hertz\tothe{0.5}}$. Additionally, we delve into analyzing communication errors using Voronoi diagrams and evaluating the communication channel capacity across different modulation schemes. We find that the maximum achievable capacity for a single communication channel equals $19.3\,\mathrm{Mbps}$ and can be achieved using the $\mathsf{QAM}$4 scheme.
\end{abstract*}

\section{Introduction}
Rydberg atoms, renowned for their exceptionally high transition electric dipole moment \cite{Saffman2010}, have garnered considerable attention in scientific circles owing to their extraordinary sensitivity of the Rydberg-Rydberg microwave transitions \cite{Borówka2024,Gordon2019,Liu2021}. Although Rydberg techniques do not beat traditional antennas \cite{Meyer2020}, they are SI-traceable and can be surprisingly broadband \cite{Zhou2022}.
They are also known for an easily reproducible detection scheme \cite{Sedlacek2012, Hall2006}, meaning that the Rydberg excitation can be reproduced in any laboratory in the world and have the same behavior.
Those distinctive properties together make Rydberg atoms a great candidate for future application in the realm of field sensors \cite{Fancher2021,Osterwalder1999}, allowing for wide detection bandwidth \cite{Cui2023}, detector miniaturization \cite{Epple2014,Paradis2019}, and even microwave-to-optical conversion \cite{Borówka2024, Han2018, Vogt2019}. Additionally, various realizations of detection schemes including microwave field modulation have led to increased sensitivity to the received fields \cite{Cai2022, Facon2016}, as well as both analog \cite{Otto2021} and digital data transmission \cite{Meyer2018, Song2019}. 

Multiple groups have recently presented Rydberg atoms-based receivers built around both rubidium \cite{Sedlacek2012, Sedlacek2013}, potassium \cite{Arias2019} and cesium \cite{Gordon2019, Holloway2019}, showing the possibility of data transmission via amplitude and frequency-modulated microwave fields \cite{Meyer2018,Anderson2021,Jiao2019}, receiving modulated signals over the wide tunable frequency range \cite{Song2019}, and multi-band communication \cite{Meyer2023, Zou2020,Holloway2021}. In this paper, we present an experimental realization of a receiver allowing for receiving modulated microwave fields at the frequencies near the 802.11b Wi-Fi standard frequency bands (2.4 GHz) \cite{IEEE_norm} using quadrature amplitude modulation scheme \cite{Chung1993}. A similar setup used to detect phase-modulated signals as well as its characterization has already been shown before \cite{Holloway2019}. We build on that concept by utilizing commercially available router antennas and maximizing the single communication channel capacity achieving record values. Additionally, we present a non-standard way of resolving sent data via the Voronoi diagrams \cite{Samuel2007}, which yield a better accuracy in error calculation for the distorted geometric structures of received symbols \cite{Herzberg1994}. We also characterize our receiver and discuss the limitations imposed onto the communication channel due to the atomic response varying with the signal frequency and amplitude of the microwave field.

\section{Experimental setup}
\begin{figure}
  \centering
  \includegraphics[width=11cm, angle=0]{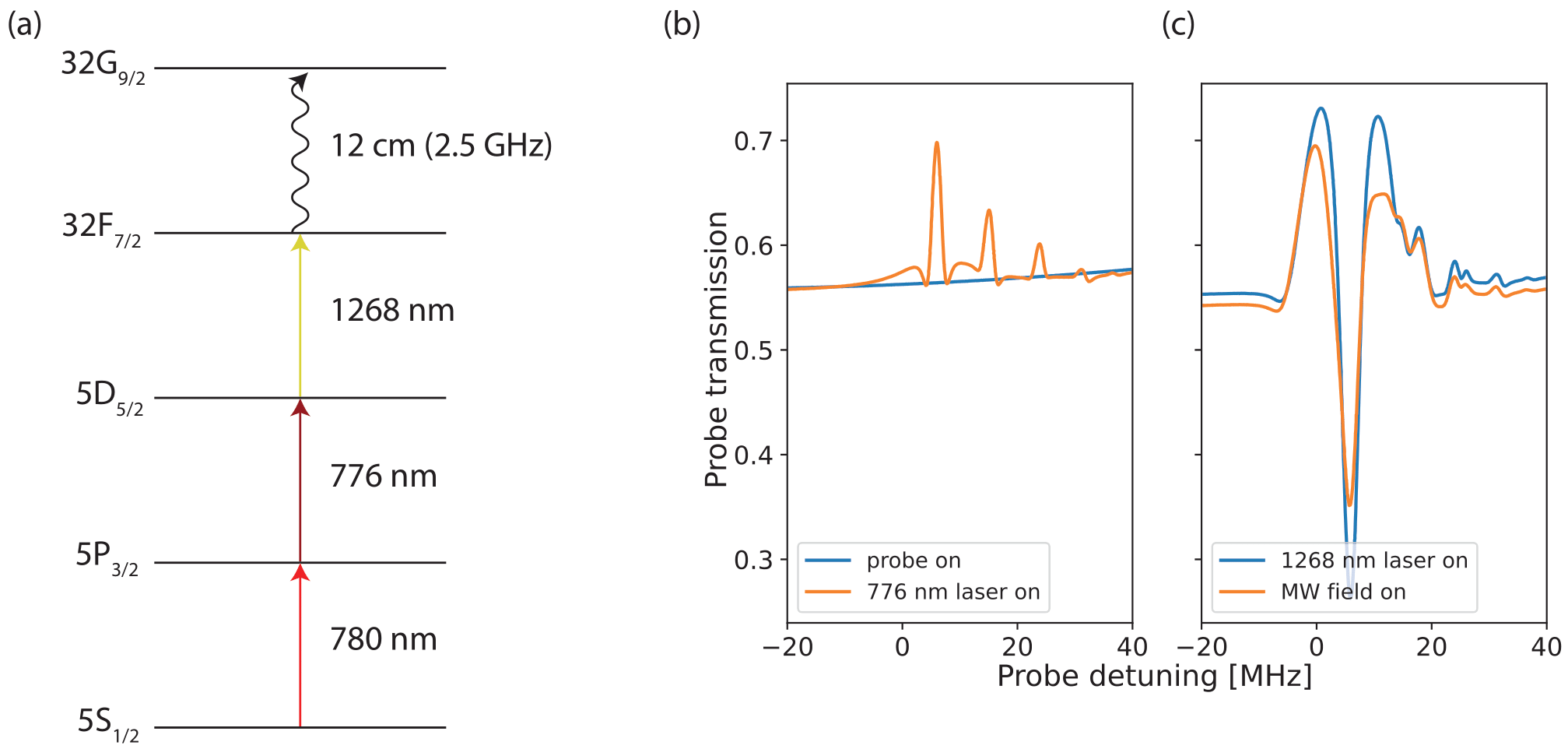}
\caption{(a) $^{85}\mathrm{Rb}$ energy level configuration considered in the experiment. (b) The blue curve represents the measured probe transmission spectrum with other fields turned off, the orange curve is the probe transmission spectrum after turning on the 776 nm laser with the EIT effect emerging. (c) The blue curve represents the probe transmission spectrum after turning on the 776 and 1268 nm lasers causing the EIA effect, and the orange curve is the probe transmission after additionally turning on the $\mathsf{MW}$ (microwave) field in the resonance with the EIA.}
\label{fig:eit}
\end{figure}

In the experiment, we consider a 5-level energy ladder of $^{85}\mathrm{Rb}$ depicted in the Fig \ref{fig:eit}a. In the following setup, the probe laser is tuned to the $\mathrm{D}_2$ transition between ground state $5^2\mathrm{S}_{1/2}(F=3)$ and $5^2\mathrm{P}_{3/2}(F=4)$. The second and third fields coupled to the $5^2\mathrm{P}_{3/2}(F=4)\rightarrow5^2\mathrm{D}_{5/2}(F=5)$ and $5^2\mathrm{D}_{5/2}(F=5)\rightarrow32^2\mathrm{F}_{7/2}$ transitions respectively. A similar atomic level configuration was used in previous works \cite{Thaicharoen2019,Brown2023}. The fields excite atoms to the Rydberg state and cause the electromagnetically induced absorption (EIA) effect \cite{Carr:12} to emerge, which can be interpreted as the interference of electromagnetically induced transparencies (EIT) \cite{Marangos1998,Fleischhauer2005} caused by the 776 and 1268 nm lasers. The last $32^2\mathrm{F}_{7/2}\rightarrow32^2\mathrm{G}_{9/2}$ transition is in the considered $\mathsf{MW}$ (microwave) regime ($f_{\mathsf{MW}} = \SI{2.5}{\GHz}$). Applying the $\mathsf{MW}$ field tuned to this transition further increases the EIA effect. The impact of the subsequent fields on the probe transmission spectrum is shown in Fig \ref{fig:eit}b and Fig \ref{fig:eit}c. It is worth noting that the transition is close enough to the 13 (spanning 2.461-2.483 GHz frequency range) channel of Wi-Fi 2.4 GHz standard that transmission using that channel can be detected. However, as the Wi-Fi channel is detuned by about $\SI{20}{\MHz}$ from the atomic resonance it significantly lowers the power level of the measured signal. As we want to maximize the data transmission through our setup we simulate the single communication channel Wi-Fi transmission at the atomic resonance frequency.
\begin{figure}
  \centering
  \includegraphics[width=11cm]{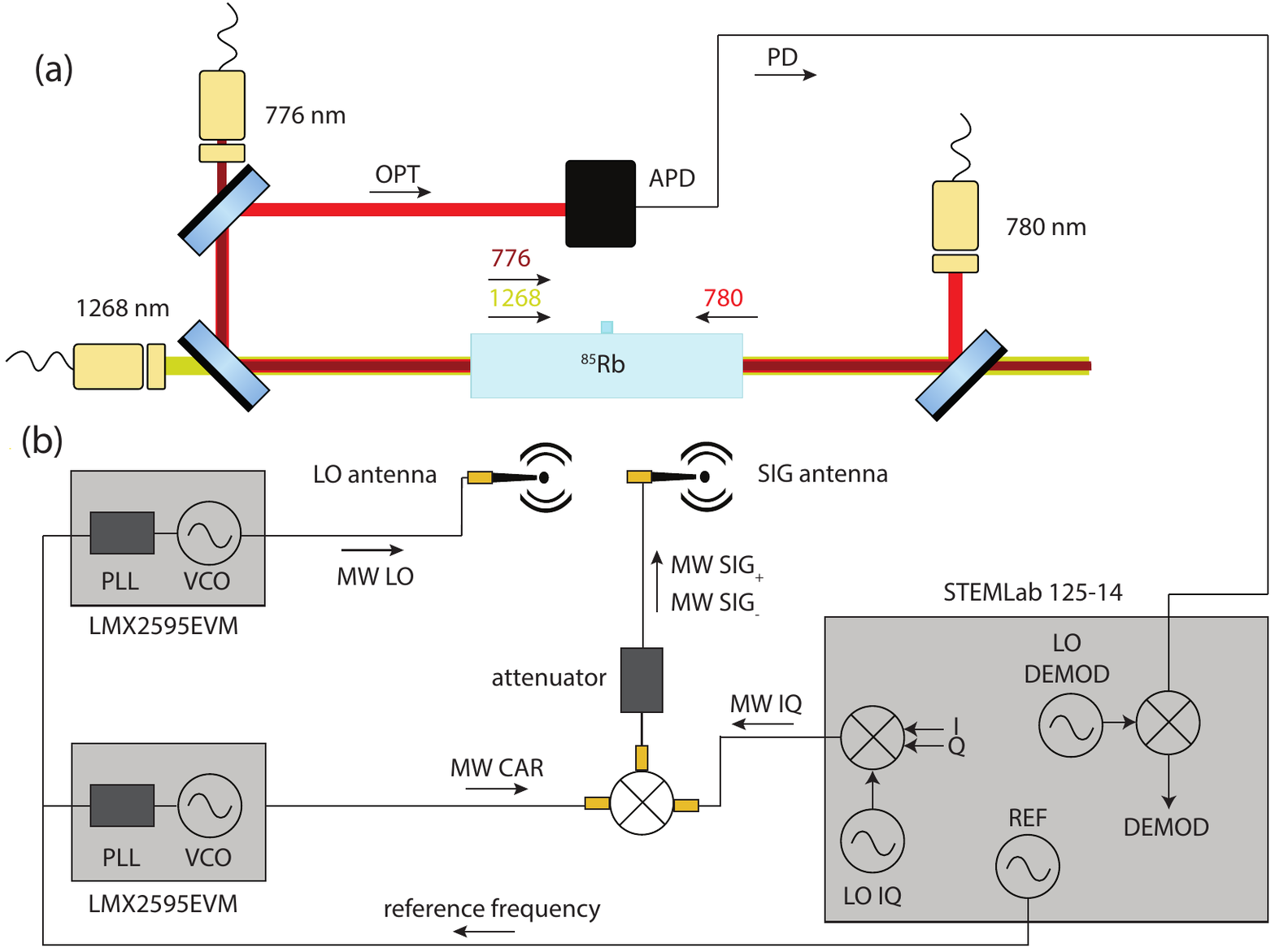}
\caption{(a) Optical part of the experimental setup. $\mathsf{APD}$ - avalanche photodiode, $\mathsf{OPT}$ - optical signal, $\mathsf{PD}$ - photodiode signal (b) The electronic part of the experimental setup. $\mathsf{PLL}$ - phase-locked loop, $\mathsf{VCO}$ – voltage-controlled oscillator, $\mathsf{LO}$ - local oscillator, $\mathsf{MW}$ - microwave, $\mathsf{CAR}$ - carrier signal, $\mathsf{DEMOD}$ - demodulation signal.}
\label{fig:scheme}
\end{figure}

Our experimental setup is built around room-temperature ($\SI{22.5}{\celsius}$) $^{85}\mathrm{Rb}$ atoms and its simplified scheme is depicted in Fig \ref{fig:scheme}. The optical part of the setup is shown in Fig \ref{fig:scheme}a. To partially reduce the thermal broadening the 776 and 1268 nm and the probe beams are counter-propagating. In terms of polarization, all beams are circularly polarized, where the probe beam is left-handed circularly polarized, and coupling beams are right-handed circularly polarized. All beams focused inside the cylindrical vapor cell with a length equal to 58 mm. The powers of lasers are chosen to maximize the EIA effect, which is achieved for probe laser power equal to $P_{780} =\SI{1.4\pm0.1}{\micro\watt}$, and powers of 776 and 1268 nm lasers equal to $P_{776} = \SI{1.5\pm0.1}{\milli\watt}$ and $P_{1268} = \SI{100\pm 2}{\milli\watt}$ and all laser beam waists equal to $w = \SI{300\pm15}{\micro\meter}$. All of the lasers are frequency-stabilized to the master laser via the cavity setups.

The $\mathsf{MW}$ generation part of the setup is depicted in Fig \ref{fig:scheme}b. $\mathsf{MW}$ dipole antennas generate linearly polarized fields at frequencies around $2.5\,\mathrm{GHz}$. To show setup feasibility as the Wi-Fi signal detector we utilize standard commercially available router antennas. In the experiment, we want to detect and we want to detect and receive data through the modulated signal ($\mathsf{SIG}$) $\mathsf{MW}$ field. We utilize the STEMLab 125-14 multipurpose tool to make the $\mathsf{IQ}$ signal. Two arbitrary signal generators inside it allow us to generate I and Q components, which are mixed with the $\mathsf{IQ}$ local oscillator ($\mathsf{LO\,IQ}$) constant waveform (CW) signal at the frequency $f_{\mathsf{IQ}}$, and the combined signal is then sent to the external frequency mixer and mixed with the carrier wave ($\mathsf{CAR}$) signal. The $\mathsf{CAR}$ signal for the $\mathsf{SIG}$ field is generated using the LMX 2595 phase-locked loop frequency synthesizer with the frequency $f_{\mathsf{CAR}}$. The signals from the synthesizer and the STEMLab 125-14 tool are then mixed via the external frequency mixer, giving rise to two sidebands at the frequencies $f_{\mathsf{SIG}\pm} = f_{\mathsf{CAR}}\pm f_{\mathsf{IQ}}$. The generated sidebands are then sent to the $\mathsf{SIG}$ antenna. 

To perform the heterodyne detection we introduce an additional $\mathsf{MW}$ field acting as a local oscillator ($\mathsf{LO}$) which is generated directly by another LMX2595 frequency synthesizer with the frequency $f_{\mathsf{LO}}$ is directly sent to the $\mathsf{LO}$ antenna. To achieve the phase-sensitive detection we utilize the fact, that atoms interacting with both $\mathsf{MW}$ fields behave as a frequency mixer \cite{Simons2019} making the signal at an intermediate frequency visible in the probe transmission, which we further refer to as optical signal ($\mathsf{OPT}$). The probe transmission is measured using an avalanche photodiode with a 50 MHz bandwidth. The photodiode signal ($\mathsf{PD}$) is then digitally demodulated at the center frequency of the $\mathsf{OPT}$ signal and the demodulated signal ($\mathsf{DEMOD}$) is retrieved. To optimize the atomic response to the $\mathsf{MW}$ fields in the heterodyne detection we shift the EIT caused by 776 and 1268 nm lasers by 3 MHz to the higher probe frequencies and lock the probe laser to the D2 transition.

\begin{figure}
  \centering
  \includegraphics[width=13cm]{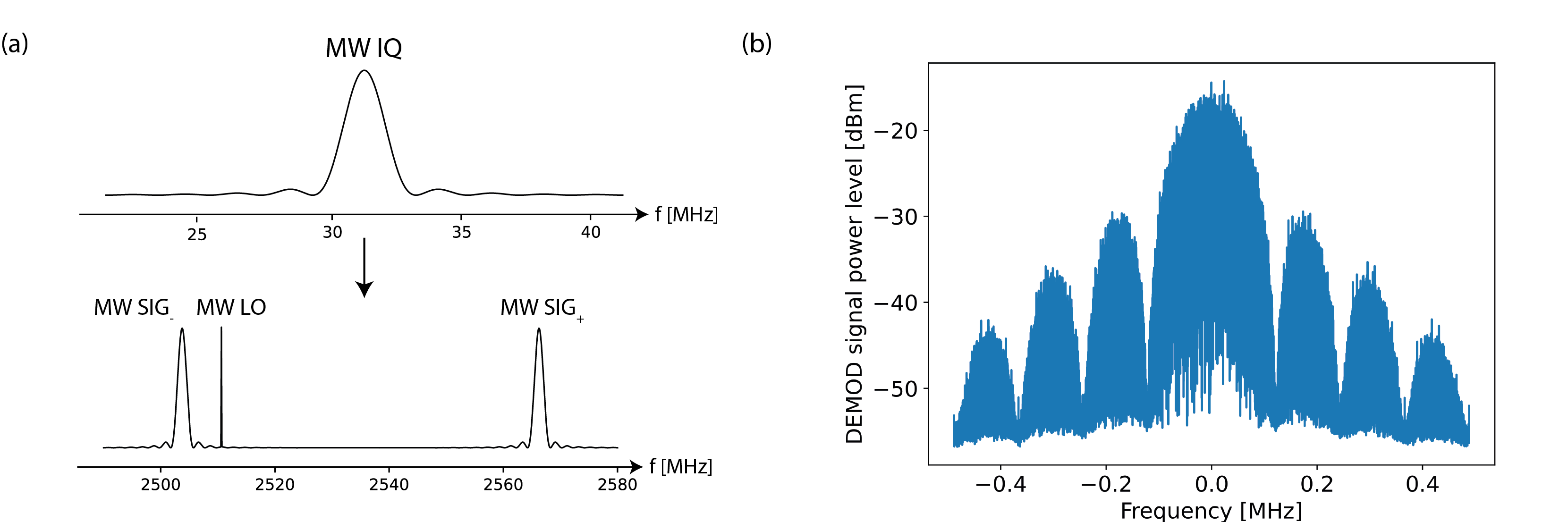}
\caption{(a) Spectra of the exemplary $\mathsf{IQ}$ signal used in the data transmission (upper spectrum) generated using STEMLab 125-14 multipurpose tool, and signals sent from both antennas (lower spectrum). The $\mathsf{SIG}$ field is prepared by mixing $\mathsf{IQ}$ signal with $\mathsf{CAR}$ sent from LMX2595 frequency synthesizer, whereas the $\mathsf{LO}$ is directly sent from the LMX2595. (b) Exemplary spectrum of the $\mathsf{DEMOD}$ signal in case of receiving quadrature amplitude modulated signal.}
\label{fig:spectras}
\end{figure}

The frequencies of the $\mathsf{MW}$ fields are chosen in such a way, that only the beat note between one of the sidebands of the $\mathsf{SIG}$ field and $\mathsf{LO}$ can be detected and the amplitude of the measured signal is optimized. For the $\mathsf{SIG}$ field, we chose $f_{\mathsf{CAR}} = \SI{2535}{\MHz}$ and $f_{\mathsf{IQ}} = \SI{31.25}{\MHz}$, so that lower sideband with the frequency of $f_{\mathsf{SIG}-} = \SI{2503.75}{\MHz}$ is in the resonance with the EIA generated by laser beams and the higher sideband with the frequency of $f_{\mathsf{SIG}+} = \SI{2566.25}{\MHz}$ is too far detuned from the EIA resonance to interact with it. The frequency of the $\mathsf{LO}$ field is set to $f_{\mathsf{LO}} = \SI{2510.59}{\MHz}$ making the $\mathsf{OPT}$ signal visible in the probe transmission with the central frequency $f_{\mathsf{OPT}} = f_{\mathsf{DEMOD}} = \SI{6.84}{\MHz}$. Local oscillator at a similar frequency was used by \cite{Elgee2023}. The spectra of the exemplary $\mathsf{IQ}$ signal and the combined $\mathsf{SIG}$ and $\mathsf{LO}$ fields are shown in Fig \ref{fig:spectras}a, and the spectrum of the corresponding $\mathsf{DEMOD}$ signal is shown in Fig \ref{fig:spectras}b. The Rabi frequencies of both fields are $\Omega_{\mathsf{SIG}} = 2\pi\cdot\SI{2.54\pm0.06}{\MHz}$ and $\Omega_{\mathsf{LO}} = 2\pi\cdot\SI{8.1\pm0.5}{\MHz}$, which corresponds to the amplitude of the electric fields $A_{\mathsf{SIG}} = \SI{5.16\pm0.12}{\milli\volt\per\cm}$ and $A_{\mathsf{LO}} =\SI{16\pm 1}{\milli\volt\per\cm}$.
\section{Results}

\subsection{Calibration of the microwave field}

\begin{figure}
  \centering
  \includegraphics[width=11cm]{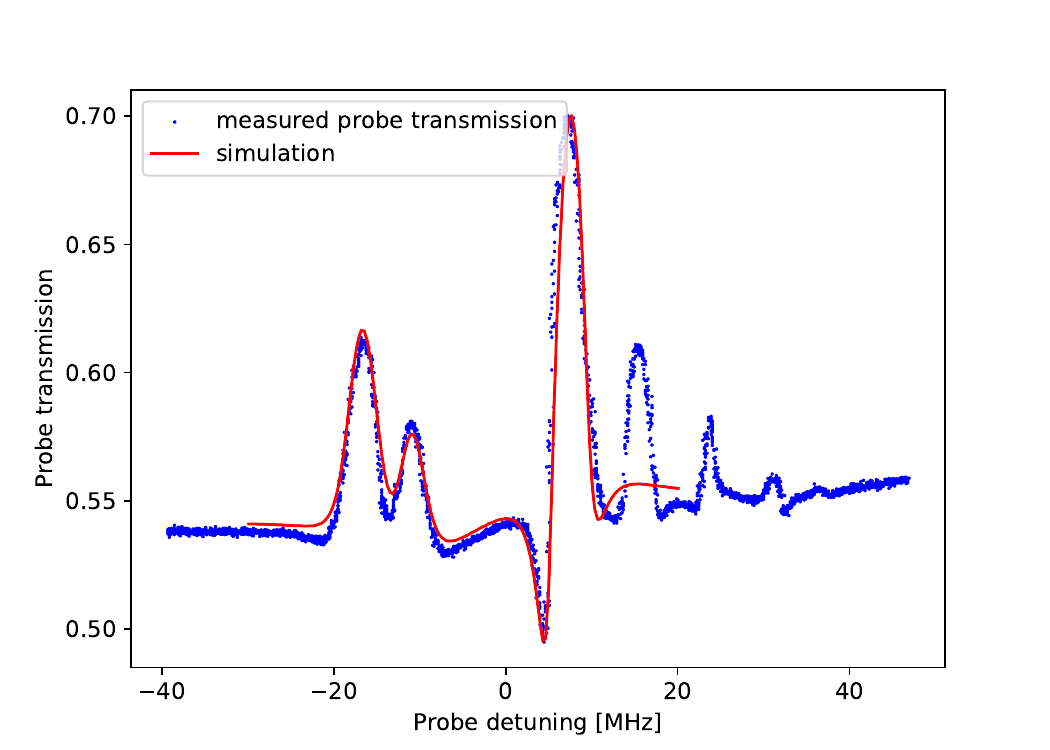}
\caption{Probe transmission spectrum (blue dots) in the case of the absolute calibration of the $\mathsf{SIG}$ field. The EIT caused by the 1268 nm laser is shifted by 22 MHz to the lower probe frequencies, so it does not interact with the EIT caused by the 776 nm laser. The red curve represents the fitted probe transmission spectrum calculated by numerically finding the steady state of the Lindblad master equation. The mismatch between the simulation and measured spectrum for the higher probe frequencies is caused by considering in the simulation only the strongest atomic transitions, which is dictated by the computational time and insignificant impact of additional atomic transitions on the behaviour of the EIT caused by the 1268 nm laser.}
\label{fig:eit_detuned}
\end{figure}

To find our receiver's sensitivity and noise level, we need to calibrate its response to the $\mathsf{MW}$ fields. For the calibration, we consider the measurements of the $\mathsf{PD}$ signal before the demodulation. We detune the 1268 nm laser by $22\,\mathrm{MHz}$ to the lower probe frequencies, so the EIT caused by the 776 and 1268 nm laser do not overlap as shown in Fig \ref{fig:eit_detuned}. We then measure the probe transmission spectrum with the EIT splitting caused only by the CW $\mathsf{SIG}$ field with the $\mathsf{LO}$ field turned off. We measure the highest amplitude of the $\mathsf{IQ}$ signal being 2 V and the signal attenuation from 0 to 15 dB with step 5 dB. The attenuator is precise up to $\pm 0.5$ dB and we use it to change $\mathsf{SIG}$ field power in a controllable manner. As a reference, we also gather data for EIT with both antennas turned off. To find the Rabi frequency for different powers of the $\mathsf{SIG}$ field we fit the numerical predictions of the probe transmission spectrum to the measured data. An example of such numerical fit calculated by finding the steady state of the Lindblad master equation compared to the experimental data is shown in Fig \ref{fig:eit_detuned}. Based on the fitted values we can also find the amplitude of the electric based on the formula $A = \frac{\hbar\Omega}{d}$, where $A$ is the amplitude of the electric field, $\Omega$ is the Rabi frequency, and $d$ is the dipole moment of the $\mathsf{MW}$ transition, which is find using Alkali Rydberg Calculator \cite{Robertson2021} and for considered transition is equal to $d=384a_0e$, where $a_0$ is the Bohr radius and $e$ is the electron charge. To get the absolute calibration line, we fit the linear function with a slope value fixed to $-1$ to the amplitude of the electric field as a function of the signal attenuation.

As absolute calibration is possible only for high enough powers of the $\mathsf{MW}$ field, we cannot calculate the sensitivity of the receiver directly from it. As for weaker $\mathsf{MW}$ fields, the splitting cannot be seen in the probe transmission spectrum. To overcome that problem, we perform the heterodyne detection for the CW signal. We then calculate the Fourier power spectrum of the $\mathsf{PD}$ signal acquired over $\SI{65}{\us}$ for the subsequent 100 waveforms. The $\mathsf{PD}$ signal power level is then calculated by finding the maximum value of the averaged Fourier power spectrum in the frequency range of about 1 MHz around $f_{\mathsf{OPT}}$. We repeat the measurement for signal attenuation ranging from 0 to 30 dB with the step of 5 dB and for the $\mathsf{IQ}$ signal amplitudes of 2, 0.5, 0.2, and 0.03 V, allowing us to map the whole atomic response range, from saturation to the noise level. We also calculate the noise power level by measuring the noise power spectrum and averaging it over the same frequency range as mentioned previously.

For the calculated $\mathsf{PD}$ signal power level we consider the logarithmic scale, in which the power level changes linearly with the attenuation. For each of the $\mathsf{IQ}$ signal amplitudes, we fit the linear function with the slope value fixed to -1. Points for the $\mathsf{IQ}$ signal amplitude of 2 V for the heterodyne detection are shifted to lie on the same line as points from the absolute calibration for the same $\mathsf{IQ}$ signal amplitude, allowing us to calculate the amplitude of the electric field based on the $\mathsf{PD}$ signal power level for the heterodyne detection. Fitted lines are then shifted to continue the absolute calibration line,
spanning the whole range of the atomic response from saturation down to noise.

\begin{figure}
  \centering
  \includegraphics[width=11cm]{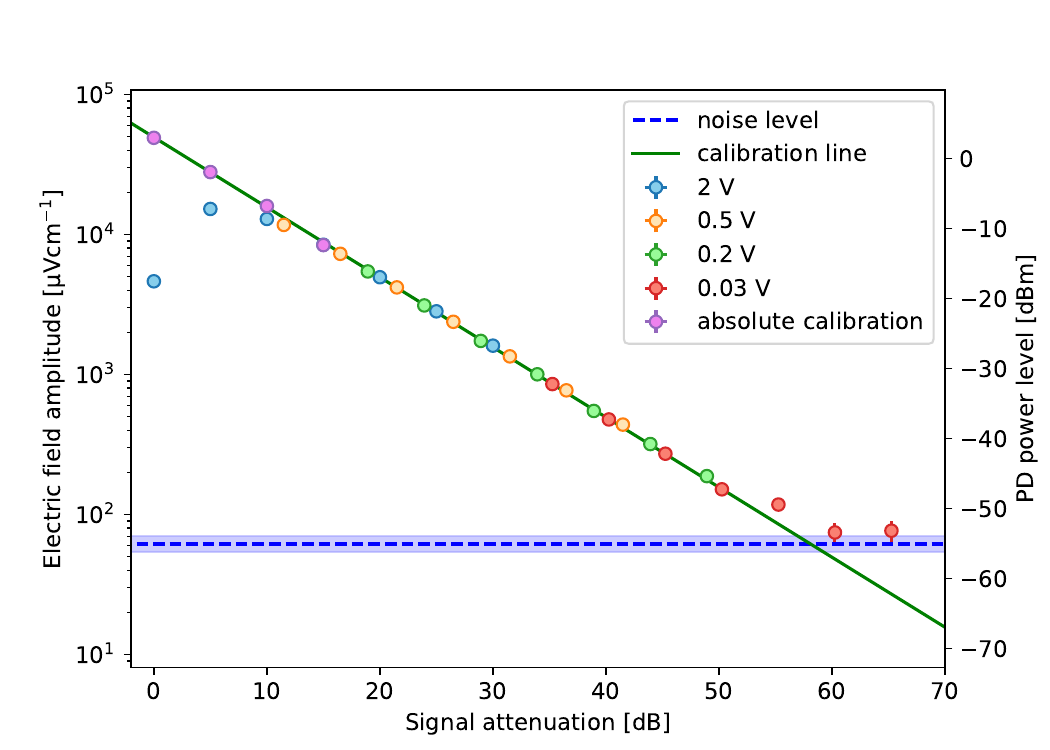}
\caption{Amplitude of the $\mathsf{SIG}$ field (left vertical axis) as the function of signal attenuation. The right vertical axis refers to the $\mathsf{PD}$ signal power level calculated from the Fourier power spectrum for heterodyne detection. The data points have been gathered during the absolute calibration and the heterodyne detection. The blue area around the shot noise level corresponds to the standard error of the calculated value.}
\label{fig:calib}
\end{figure}

It is seen in Fig \ref{fig:calib} that some points for calculated electric field amplitude lie in the linear atomic response regime of the heterodyne detection for the CW signal, based on them we find a relation between electric field amplitude and the $\mathsf{PD}$ signal power level. Using the found relation, we calculate the amplitude of the electric field for which the $\mathsf{PD}$ signal power level is equal to the noise power level in the Fourier power spectrum. The electric field amplitude corresponding to that signal level is equal to $\SI{62\, \pm 8}{\micro\volt\per\cm}$ which in case of the $\SI{65}{\us}$ measurement time window translates to a noise level equal to $\SI{0.50\pm0.06}{\micro\volt\per\cm\per\hertz\tothe{0.5}}$. The relation between the electric field amplitude and the $\mathsf{PD}$ signal attenuation together with the calibration line are shown in Fig \ref{fig:calib}.
We also find the amplitude of the $\mathsf{SIG}$ field corresponding to the saturation level, which we define as a power level 1 dB weaker than the power for which the change in the atomic response due to changing signal attenuation becomes nonlinear. The calculated value of the amplitude of the electric field corresponding to the saturation level is equal to $A_{sat} = \SI{6.5\pm0.1}{\milli\volt\per\cm}$.

\subsection{Atomic response range}
\begin{figure}
  \centering
  \includegraphics[width=11cm]{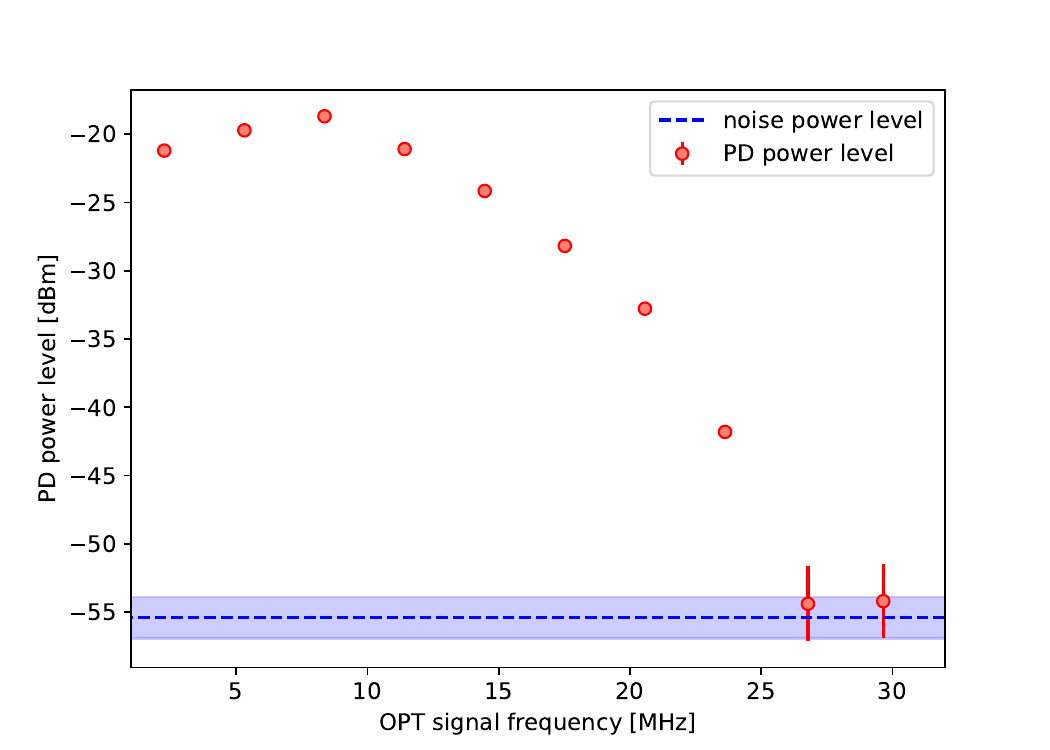}
\caption{Atomic response to the  $\mathsf{MW}$ field in the case of heterodyne detection as a function of frequency of the $\mathsf{OPT}$ signal. The response was found by sending the frequency comb with spacing between components equal to $\SI{3.04}{\MHz}$ and then, based on the Fourier power spectrum, finding the measured power level of each of the frequency components. The blue area around the line representing the shot noise power level corresponds to the error of the value of the noise power level.}
\label{fig:bandwidth}
\end{figure}
To further characterize the capabilities of our receiver we measure its frequency response range. The measurement is performed by sending the frequency comb with the frequency spacing equal to $\SI{3.04}{\MHz}$ and the maximum amplitude of the electric field equal to $A = \SI{4.06\pm0.12}{\milli\volt\per\cm}$. The frequency comb is generated using the STEMLab 125-14 multipurpose tool as a linear combination of the sines waves with the frequencies corresponding to each of the comb components and the central frequency equal to $f_{IQ}$. The frequency comb is then sent to the external mixer where it is mixed with the $\mathsf{CAR}$ signal. We measure 100 consecutive waveforms of $\mathsf{PD}$ signal, each lasting $\SI{65}{\us}$. For each of the waveforms, we calculate the Fourier power spectrum of the $\mathsf{PD}$ signal for the heterodyne detection. The $\mathsf{PD}$ signal power level of each of the frequency components is found by taking the maximum value of the spectrum in the frequency range of 1 MHz around the expected frequency of each component. To find the noise power level we again collect 100 waveforms lasting $\SI{65}{\us}$ with both antennas turned off. The noise power level from a single measurement for each of the frequency components is found by averaging the Fourier power spectrum of the $\mathsf{PD}$ signal over the same frequency range in which the power level of the component has been found. The averaged $\mathsf{PD}$ power signal level as a function of the frequency of the $\mathsf{OPT}$ signal is shown in Fig \ref{fig:bandwidth}, together with the shot-noise power level derived by averaging the noise power level over the whole spectrum and each of the measurements. The visible drop in the measured $\mathsf{PD}$ signal power level for higher $\mathsf{OPT}$ signal frequencies can be attributed to the decreasing atomic response limiting available bandwidth. Based on estimated $\mathsf{PD}$ signal and noise power levels, we calculate the $\mathsf{SNR} = S/N$, where $S$ is the $\mathsf{PD}$ signal power level and $N$ is the noise power level. Both values are found by averaging calculated signal and noise power level over each of the measurements.

Knowing the $\mathsf{SNR}$ of the $\mathsf{PD}$ signal at frequency $f_{\mathsf{OPT}}$, one can determine the capacity of the communication channel at this frequency, which serves as a tight upper bound on the data transmission bitrate achievable with a given setup. Because the heterodyne detection noise is essentially Gaussian, this maximum is given by the Shannon-Hartley formula~\cite{Shannon1948}:
\begin{equation}
    C(f_{\mathsf{OPT}}) = B \log_2(1+\mathsf{SNR}(f_{\mathsf{OPT}})).
\end{equation}
Here, $\mathsf{SNR}(f_{\mathsf{OPT}})$ is the signal-to-noise ratio determined for a given frequency of the $\mathsf{OPT}$ signal, whereas $B$ is the bandwidth allocated to the communication channel, which in our case is equivalent to the $\SI{3.04}{\MHz}$ spacing between subsequent values of $f_{\mathsf{OPT}}$. Each of such $B$-wide channels at the subsequent $f_{\mathsf{OPT}}$ values can be used simultaneously -- a standard practice known as frequency division multiplexing (FDM) \cite{IEEE_norm} -- which gives then finally the total capacity as a sum over all the channels, $C_{\text{tot}} = \sum_{f_{\mathsf{OPT}}} C(f_{\mathsf{OPT}})$. Doing so for our setup, we find the total capacity achievable to be $C_{\text{tot}} = \SI{234.2\pm1.9}{\mbps}$.

\begin{figure}
  \centering
  \includegraphics[width=11cm]{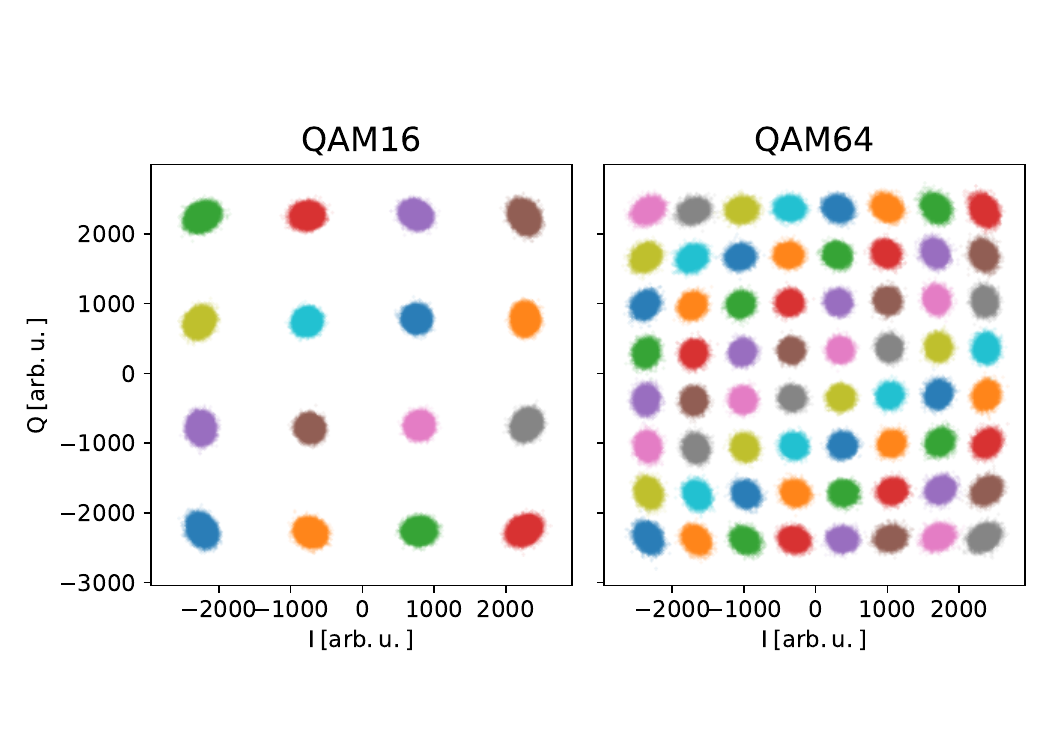}
    
    \caption{The received points for the symbol frequency $\SI{122}{\kHz}$ for the $\mathsf{QAM16}$ and $\mathsf{QAM64}$ schemes. The symbol sequences are generated randomly with the flat probability distribution of sending each of the symbols in a given $\mathsf{QAM}$ scheme. To get rid of the global phase, we additionally send a known symbol, the same for each transmission, acting as a phase reference. Each cluster of points corresponds to a unique sent symbol. }
    \label{fig:qam1664}
\end{figure}
\begin{figure}
  \centering
  \includegraphics[width=11cm]{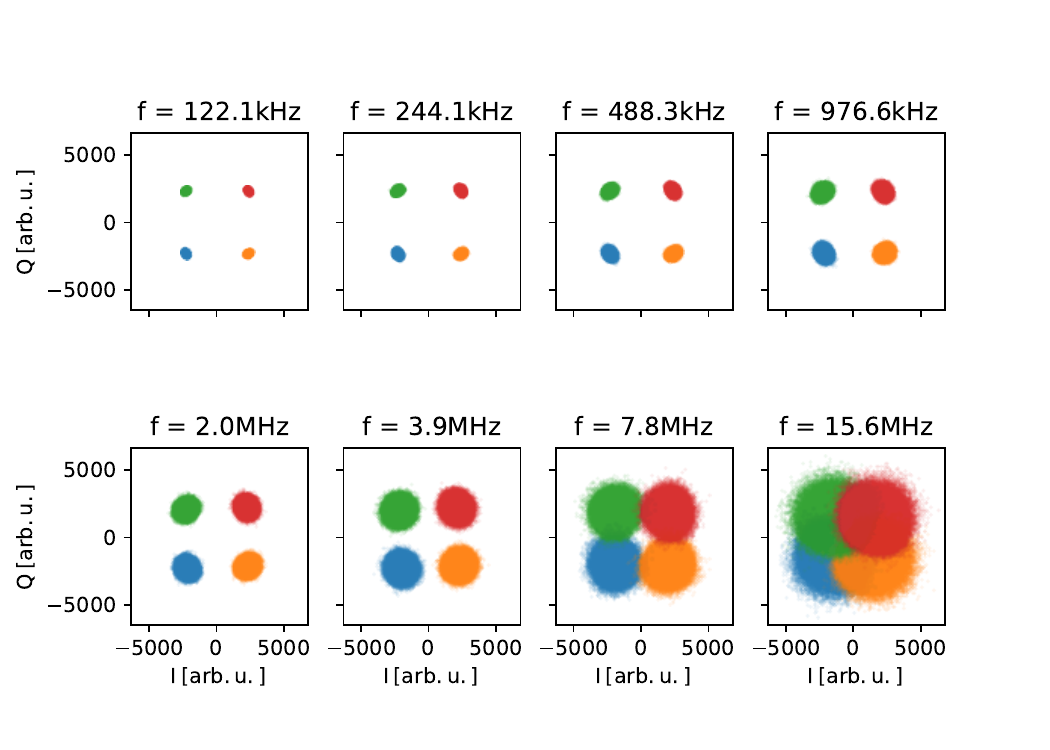}
\caption{The received points for $\mathsf{QAM4}$ scheme for different symbol frequencies. The symbol sequences are generated randomly with the flat probability distribution of sending each of the symbols in a given $\mathsf{QAM}$ scheme. To get rid of the global phase, we additionally send a known symbol, the same for each transmission, acting as a phase reference. Each cluster of points corresponds to a unique sent symbol. }
\label{fig:qam4}
\end{figure}

\subsection{Data transmission protocol}
The main goal of the experiment is to receive data encoded into the signal $\mathsf{SIG}$ through modulation of the $\mathsf{MW}$ field, using the quadrature amplitude modulation ($\mathsf{QAM}$) scheme \cite{Campopiano1962}. In this scheme, the signal is generated from the amplitude-modulated I (in-phase) and Q (quadrature) components shifted in phase by $\nicefrac{\pi}{2}$. The components define two-dimensional IQ space, in which $\mathsf{QAM}$ can be described as an even symmetrical points distribution around the origin.

We consider the signal modulation protocols using 4, 16, and 64 symbols which correspond to 2, 4, and 6 bits per symbol. Symbols are sent with the symbol frequency $f_{sym}$ ranging between $\SI{120}{\kHz}$ and $\SI{16}{\MHz}$. Due to technical limitations of our electronic setup, only specific values of $f_{sym}$ corresponding to $\nicefrac{125}{2^n}\,\mathrm{MHz}$, where $n\in[3,4,...,10]$, can be received correctly. The symbol sequences are generated randomly with the flat probability distribution of sending each of the symbols in a given $\mathsf{QAM}$ scheme. To get rid of the global phase, we additionally send a known symbol, the same for each transmission, acting as a phase reference.

We collect 100 subsequent waveforms of the $\mathsf{DEMOD}$ signal for a given symbol sequence. For each modulation scheme and symbol frequency, every sent symbol consists of the same number of data points. The measured $\mathsf{DEMOD}$ signal is divided into segments corresponding to the sent symbols, which are then averaged over their duration to retrieve values of the I and Q components of the symbol. The higher the $f_{sym}$, the fewer data points for each of the received symbols can be retrieved, as the delay caused by the atomic response and applying frequency filters during demodulation becomes more prominent. We then plot received symbols as points in the IQ phase space, which for $\mathsf{QAM16}$ and $\mathsf{QAM64}$ with $f_{sym} = \SI{122}{\kHz}$ are presented in Fig \ref{fig:qam1664}, and for the $\mathsf{QAM4}$ for different symbol frequencies are shown in Fig \ref{fig:qam4}. As the maximum amplitude of the sent symbols is near the saturation of the heterodyne detection it can be noted, that the point distribution corresponding to the highest amplitude of the signal is flattened radially, which can be attributed to the deviation of the atomic response from the linear regime. It is also worth noting that for higher powers of the signal fields the nonlinear response impacts the symbol distribution and the shape of the whole measured $\mathsf{QAM}$ constellation significantly. As the signal power increases over the saturation limit, different $\mathsf{QAM}$ symbol clouds quickly start to overlap, leading to a rapid decrease in transmission efficiency. Moreover in Fig \ref{fig:qam4}, it can be seen that as the symbol frequency increases, the noise level, corresponding to the points spread in IQ space, also increases. The increase of the shot-noise power level is caused by the change in the measurement time of the transmission sequences.

One of the more important parameters of data transmission is the symbol error rate ($\mathsf{SER}$), defined as the probability of incorrectly receiving symbols. As the noise for both the I and Q components is Gaussian, we calculate $\mathsf{SER}$ by integrating Gaussian probability distribution over a specific area of the IQ space.
The integral areas are defined by the structures known as the Voronoi cells \cite{Aurenhammer1991, Okabe2009}. The diagram is created from seed points, which in this case are mean positions received symbols, based on which the Voronoi cell limits are calculated. The Voronoi cell is defined as a set of points which are closer to a given seed than to the other seeds. Such an approach takes into account the distortions of the received signal in phase space which occurs more often for higher frequencies as well as distortions of the symbol constellations caused by saturation effects, which yields better results compared to the square region approach.

\begin{figure}
  \centering
  \includegraphics[width=11cm]{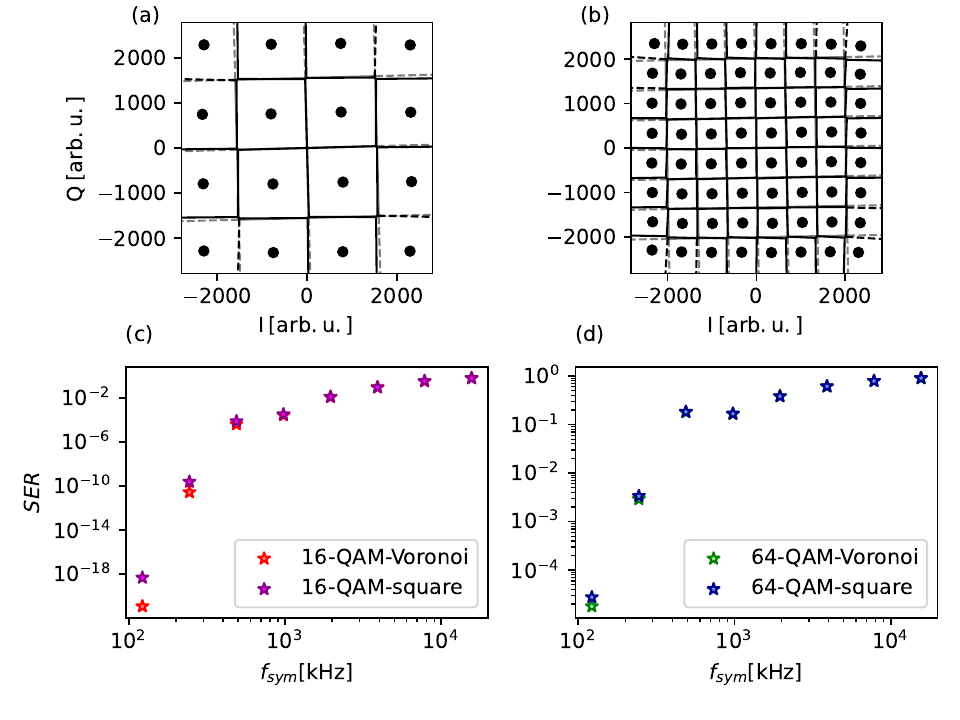}
\caption{(a-b) Voronoi diagram in the case of transmission frequency $\SI{122}{\kHz}$ for the $\mathsf{QAM16}$ (a) and $\mathsf{QAM64}$ (b) modulation. The symbol sequences are generated randomly with the flat probability distribution of sending each of the symbols in a given $\mathsf{QAM}$ scheme. To get rid of the global phase, we additionally send a known symbol, the same for each transmission, acting as a phase reference. Black points represent the mean positions of the received symbols. The grey diagram is an orthogonal square grid. Both the solid and dashed lines represent the borders of the corresponding Voronoi cell. The solid lines represent finite borders, i.e. the borders, which end by crossing with another border line, and the dashed lines represent infinite borders. 
(c-d) Comparison of errors obtained from Voronoi diagrams and an orthogonal grid for $\mathsf{QAM16}$ (c) and $\mathsf{QAM64}$ (d) as a function of symbol frequency.
}
\label{fig:vor}
\end{figure}

For a given measured symbol sequence we first calculate the mean position and standard deviation of each of the measured symbols distributions in every direction. Based on the mean positions we then generate the Voronoi diagram. The standard deviations in each direction are used for the error calculation procedure. Examples of generated Voronoi diagrams together compared to the orthogonal grids are shown in Fig \ref{fig:vor}a and Fig \ref{fig:vor}b. Both solid and dashed lines represent the borders of the Voronoi cells, with the dashed lines representing infinite borders, and solid lines representing finite borders, i.e. borders ending by crossing with another border line. To find the symbol error rate for each of the $\mathsf{QAM}$ symbols, we calculate the Gaussian probability distribution over an area outside the respective Voronoi cell. The calculated values for every $\mathsf{QAM}$ symbol are then averaged giving the average $\mathsf{SER}$. To further quantify the comparison between Voronoi diagrams and the standard square regions approach we also calculate the average $\mathsf{SER}$ for the orthogonal grid. The $\mathsf{SER}$ as a function of symbol frequency for both approaches are shown in Fig \ref{fig:vor}c for $\mathsf{QAM}$16 and Fig \ref{fig:vor}d for $\mathsf{QAM}$64. 

\begin{figure}
  \centering
  \includegraphics[width=11cm]{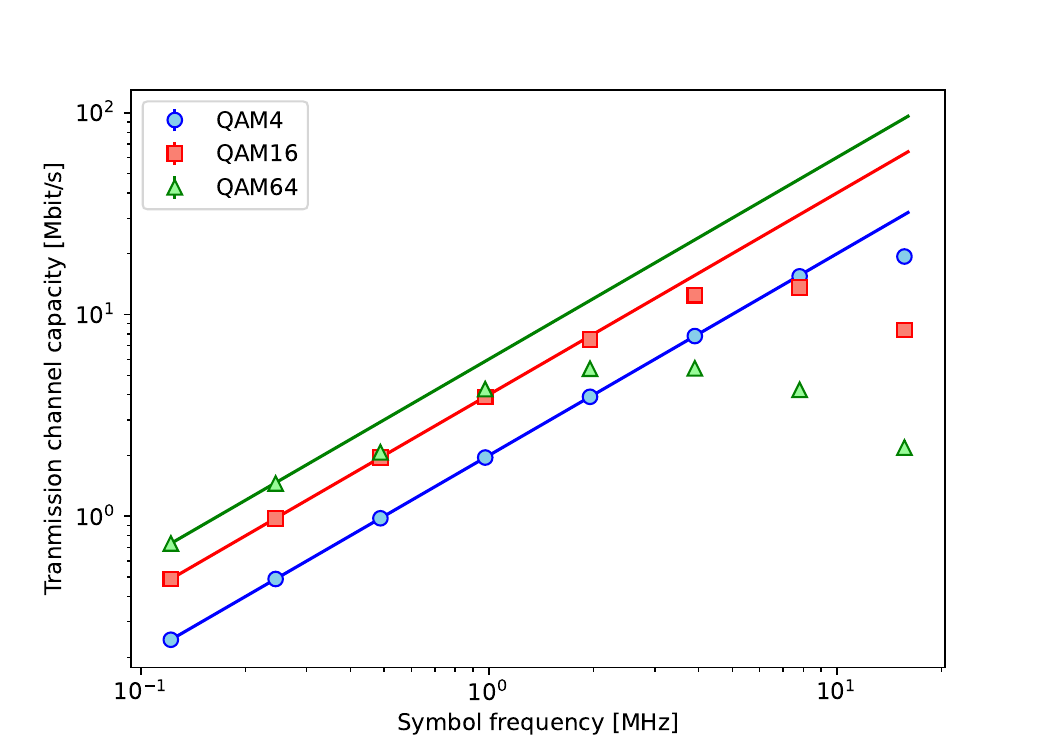}
\caption{Channel capacities for calculated $\mathsf{SER}$ for each of the measured modulation schemes as a function of symbol frequency. The $\mathsf{SER}$ was calculated by integrating Gaussian probability distribution over individual Voronoi cells in the Voronoi diagram. Calculated channel capacities correspond to a pessimistic case in which $\mathsf{SER}$ is distributed evenly between each of the $\mathsf{QAM}$ symbols.  The continuous lines represent channel capacities in case of errorless communication.}
\label{fig:capacity}
\end{figure}

Based on the calculated $\mathsf{SER}$ for each of the modulations and symbol frequencies, we find transmission channel capacity defined as the maximum data transmission speed achievable with error correction codes. To estimate it, we consider the most pessimistic scenario of the symmetrical communication channel in which $\mathsf{SER}$ is distributed evenly between the symbols. The channel capacity in that case is given as:
\begin{equation}
    C = f_{sym}\left(\log_2(M)-\mathsf{SER}\log_2(M-1)+\mathsf{SER}\log_2(\mathsf{SER})+(1-\mathsf{SER})\log_2(1-\mathsf{SER})\right)
\end{equation}
where $M$ is the number of symbols in the constellation. The channel capacity for measured points for each of the $\mathsf{QAM}$ schemes as a function of symbol frequency compared to the capacity for errorless communication is shown in Fig \ref{fig:capacity}. It can be seen that the highest capacity is achieved for the $\mathsf{QAM4}$ scheme for the $f_{sym}=\SI{15.6}{\MHz}$ with the average $\mathsf{SER} = 13\%$  and is equal to $\SI{19.33\pm0.04}{\mbps}$.
\section{Summary}
In this paper, we presented and characterized a Rydberg atom-based receiver capable of detecting $\mathsf{MW}$ signals at the frequency near the 802.11b Wi-Fi standard frequency bands \cite{IEEE_norm} with a maximum channel capacity in the pessimistic scenario equal to $19.3\,\mathrm{Mbps}$ and shot noise sensitivity equal to $\SI{0.50}{\micro\volt\per\cm\per\hertz\tothe{0.5}}$. Despite considering in the experiment only quadrature amplitude modulation schemes, other modulations, such as quadrature phase shift keying, can also be achieved in our setup. It is worth noting, that due to the nearest commonly used Wi-Fi channel being at the frequency of about $\SI{20}{\MHz}$ from the atomic transition, the efficient detection of Wi-Fi would require additional tuning of laser or DC and AC-Stark shifts of the energy levels \cite{Ma:20}. Moreover, as the antennas used in the experiment are commercially available router antennas, we believe our findings show the possibility of detecting such transmission just by defining another Wi-Fi channel with a central frequency corresponding to the atomic resonance.

The receiver's bandwidth, and thus also maximum symbol frequency, are strongly limited by the atomic response, which drops rapidly for higher frequencies of the optical signal. Such behavior was also reported by other groups \cite{Simons2019, Holloway2019}. Moreover, we found that the signal power is also limited by the saturation and noise power levels, restricting available IQ phase space, and in effect, achievable communication channel capacity.

Despite limitations, the communication channel capacity could still be increased by utilizing multiple narrowband communication channels, which could further increase the transmission channel capacity. The presented setup can also be extended to perform microwave-to-optical conversion \cite{Borówka2024,Rueda2016} allowing for further processing of the sent information in the optical setup.
\paragraph{Funding.} Narodowe Centrum Nauki (2021/43/D/ST2/03114);

\paragraph{Acknowledgments.} This research was funded in whole  or in part by National Science Centre, Poland grant No. 2021/43/D/ST2/03114. The “Quantum Optical Technologies” (MAB/2018/4) project was carried out within the International Research Agendas programme of the Foundation for Polish Science co-financed by the European Union under the European Regional Development Fund. We thank K. Banaszek for the generous support and S. Borówka for support and discussions.

\paragraph{Disclosures.} The authors declare no conflicts of interest.

\paragraph{Data availability.} Data has been deposited at Harvard Dataverse \cite{DVN/EEBFSX_2024}
\bibliography{refs}
\end{document}